# Dynamic Lockstep Processors for Applications with Functional Safety Relevance

Hans Dermot Doran
Zurich University of Applied Sciences
Institute of Embedded Systems
8401-Winterthur, Switzerland
donn@zhaw.ch

Timo Lang
Zurich University of Applied Sciences
8401-Winterthur, Switzerland
langtim1@students.zhaw.ch

*Abstract*— Lockstep processing is a recognized technique for helping to secure functional-safety relevant processing against, for instance, single upset errors that might cause faulty execution of code. Lockstepping processors does however bind processing resources in a fashion not beneficial to architectures and applications that would benefit from multi-core/-processors. We propose a novel on-demand synchronizing of cores/processors for lock-step operation featuring post-processing resource release, a concept that facilitates the implementation of modularly redundant core/processor arrays. We discuss the fundamentals of the design and some implementation notes on work achieved to date.

*Keywords—functional safety, safe processing, high availability, lockstep processors, FPGA, SoC, Multicore Processors,*

## I. Introduction

Processing on processors in functionally safe applications binds additional resources [1]. Typical solutions to detect single event upsets include utilizing redundancy by carrying out the functionally safe code twice, in parallel or in series, and then comparing the execution or the results. Series execution is inefficient in terms of latency, parallel execution in terms of cost. The most common form of parallel execution architecture, lockstep processing, features two processors executing the same code either at the same time or staggered by some small number (1..2) of clocks. This technique, commonly understood as tightly-coupled lockstepping, compares the bus activities of the processors and, generally, asserts a reset should the two differ. Whilst there is a very fast reaction to errors, within a few clock cycles, there is generally no scope for degraded operation and the monitoring circuitry may slow the execution speed of the processors. Additionally, memory and possibly I/O requires separate protection, typically achieved using ECC.

Loosely-coupled lockstep processing is generally taken to mean two processors that tick to their own clocks and use separate ROM and RAM and, generally, where results of operations are compared rather than bus activity. The increase in RAM and ROM space for duplicated storage and, potentially, slower error detection [2], is offset by the prospect of supporting software and hardware diversity, a degraded operation mode and the superfluity of ECC based data protection. Undeniably, the duplication of ROM/RAM is costly, especially in integrated circuit solutions [3].

There are numerous single-chip lockstep solutions available [4, 5], popular in the very cost-sensitive automotive industry. In other domains where the use of multicore processors is common but eschew the additional cost of a single-chip lockstep, researchers grapple with the question of how to leverage features found in multicore architectures including debug features [6], re-configurability [7] and core isolation [8], albeit most such solutions require additional loosely-coupled lockstepping to ensure safe processing. Researchers [9] also suggest scheduling non-critical tasks on processors normally reserved for critical-task execution.

We therefore propose a novel dynamic lockstepping architecture in which otherwise unrelated homogeneous cores can independently accept a request to join in lock-step to process a critical task. Once this task has been executed, the processors release themselves and are available for other tasks. This architecture proposal exhibits several advantages namely: non-permanent allocation of processing resources for critical/safe-processing tasks; potential to increase availability through MooN configurations; potential to perform degraded operation in case of error detection; potential to perform sanity checks on failed processors whilst upholding the application and re-integration of processors that pass the sanity check; much higher flexibility in the scheduling of processing resources across the entire application.

The paper is structured accordingly. In Section II we make our design proposal, we briefly mention some implementation notes in Section III and conclude in Section IV, drawing conclusions and proposing future work.

## II. Proposal

We can model the proposed system as a state machine, Figure 1. For simplicities sake we do not consider features such as degraded operation. The system begins in the boot state which performs checks and can transition into the (permanent) safe state if the checks do not succeed. If the checks do succeed the system can enter normal processing mode. If a processing block (f.i microprocessor running an application) demands safe processing then the system enters a transient state in which it is attempted to synchronise enough processing blocks to achieve

We acknowledge the sponsorship of this project by the Swiss Commission for Technology and Innovation (CTI), number 177277.1PFES-ES with gratitude.

the required MooN configuration, such as 2oo3 or a 3oo5. If the required configuration can be achieved the system then transitions into the safe-processing mode. In this mode N-M processing blocks may fail before the system transitions into the safe state. Alternatively the task may complete correctly and the system transitions back into the normal processing mode.

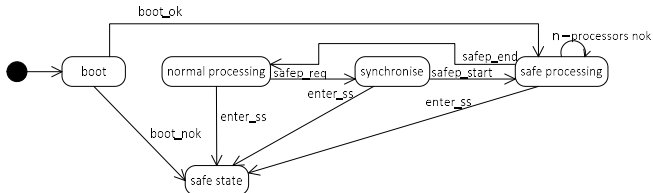

Figure 1: State Machine Proposed System

At a behavioural level we can illustrate the components using the following activity diagram (Figure 2). Initially there are, in the normal processing state, different tasks (`app_1 … app_n`) executing asynchronously on several processing blocks. Some signal must be generated to transition the sub-system into the safe processing state. If N processing blocks are required then some unit, in this case the `lock_step_monitor`, must ensure that N of these processing blocks are properly synchronised after some time period, if not then the system must transition into the safe state. If the processing blocks can be synchronised then the sub-system executes `safe_app`, monitored by the `lock_step_monitor`. If the system does not complete properly, then the system must transition into the safe state, otherwise the execution of (`app_1 … app_n`) may resume.

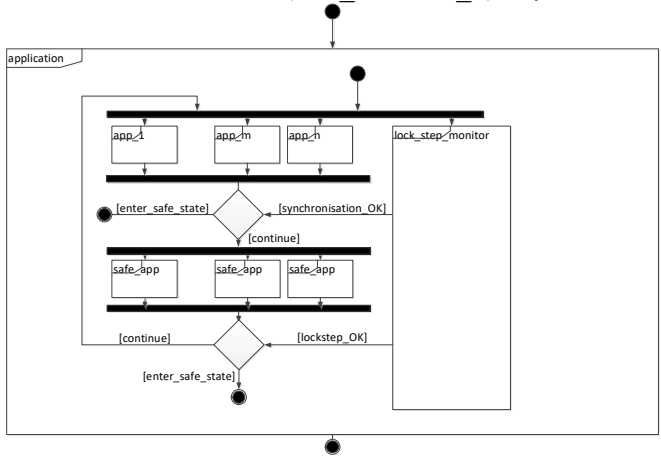

Figure 2: Activity Diagram: Application

We consider the entry mechanisms to the safe processing state, using Figure 3 which depicts the HW component `processing_block`. We allow for the use of a monitor, operating system or other such infrastructure software, which we simply term `monitor`, as well as the application. Requests for safe processing (`request_sp`) may come from this monitor, timed or triggered. Alternatively the application may request a safe processing state, again either timed or triggered. We must also allow for sources external to the processing block to request entry into the safe processing state for instance HW interrupts normally in the scope of the application or normally outside of the scope of the application. When triggered, the processing block transitions into a nominally safe mode – whilst lockstep processing is not available the code is simple enough to be inspected – asserts a `request_sync` signal and waits, for instance, by idling on a bus transfer that is prevented from completing. When some unit asserts the `continue` signal then the safe application code can proceed albeit, and unknown to the `processing_block`, in lockstep mode. When the safe code completes, the state transitions to normal processing and the `processing_block` may expect to return to whatever code it was processing before safe state was requested. We envisage an interrupt service routine (ISR) as the simplest basis for the `safe_app`.

This design translates into 16 functional requirements, [R1 … R6] for the processing block, [R7 … R14] for the lock_step_monitor and R15, R16 for any external entity. The relationship between the requirements is illustrated in (Figure 4.)

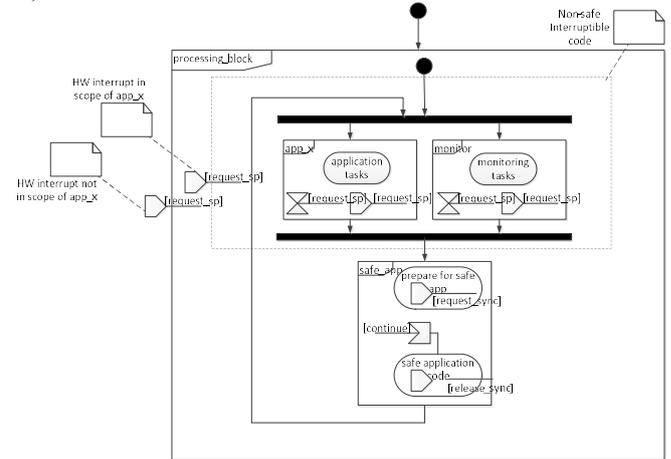

Figure 3: Activity Diagram Processing Block and Sources of Synchronization Requests

The sequence diagram below (Figure 5) illustrates a possible sequence of events with an example of two cores (`processing_blocks`) required for a 2oo2 configuration. `core_1` triggers the safe-processing by issuing a trigger signal (`trigger_sp`.) This causes the `lock_step_monitor` to generate a request signal to all attached cores to which `core_2` and `core_n` respond instantaneously. These issue a `enter_sp` bus transfer which at first stalls. As soon as the `lock_step_monitor` has received two participation requests it releases the stalled bus transfers and safe processing begins on `core_2` and `core_n`. `core_1`'s request arrives later and is rejected. After `core_2` and `core_n` both issue an `exit_sp` bus transfer, safe-processing ends. Note that despite requesting lockstep processing, `core_1` does not participate in it.

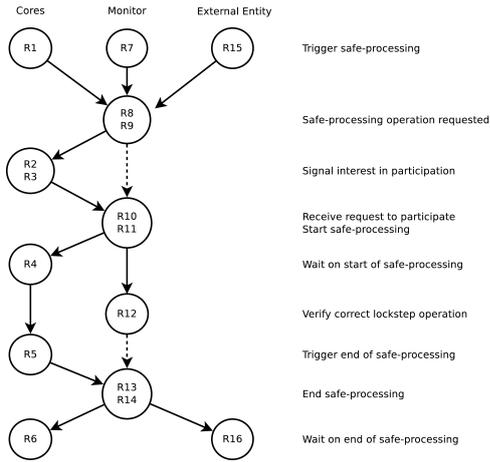

Figure 4: Control Flow and Component Requirements in the Dynamic Lockstep System

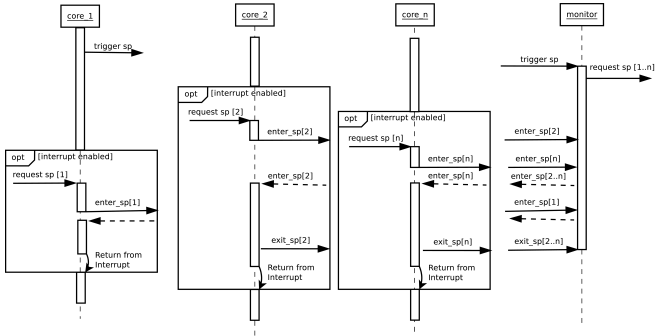

Figure 5: Sequence Diagram Detailing Lockstep Processing Request

## III. Implementation

While this proposal is of particular interest for integration on integrated circuit multicores, a cost efficient prototype implementation is easily possible using soft-cores on an FPGA, for which the two major suppliers offer IEC 61508 certified design flows [10, 11]. Intel represents an attractive solution because the Avalon bus is quite simple and is implemented as a direct connect from processor to device.

Our HW architecture (Figure 6) consists of a number (three) of `processing_blocks` each requiring an Avalon interface (Av[1..3]). If each is configured to access `system_RAM` then, during synthesis, `system_RAM` would have an arbiter attached and the arbiter would offer a port to each `processing_block` and arbitrate between simultaneous accesses. Similarly, our `lock_step_monitor` offers three ports, one for each `processing_block`. The `lock_step_monitor` also requires, a separate RAM for code and data and, optionally, an input/output device. The code to be executed safely must be loaded into `ls_RAM` during system initialisation. This `ls_RAM` could, be for instance, triple modular redundant RAM, as often encountered in safe systems.

The sequence of operation functions as shown in the sequence diagram can be visualised by the NIOS II ISR assembler code in Figure 7.

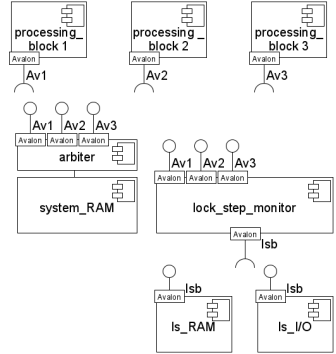

Figure 6: Architecture of System using NIOS II and the Avalon Bus

For simplicities sake we implemented a two-step procedure, the first to marshal the correct number of `processing_blocks` to perform lockstep processing and the second to actually synchronise the `processing_blocks` and lockstep through the safe code. This is reflected in the code. The `processing_block` signals interest in participating in lockstep processing by initiating a read transaction to a specific address (LOCKSTEP_SYNC_ADDRESS.) The address read will return either an accept or a reject. If participation is accepted the ISR will call a safe sub-routine, stored at SAFECODE_START. Once lock-step processing is to stop, a second read of LOCKSTEP_SYNC_ADDRESS must be performed to ensure a controlled release out of lock-stepping.

```
1   movia r2,  LOCKSTEP_SYNC_ADDRESS
2   ldbuio r2,0(r2) # issue read transfer (blocking)
3   andi r2,r2,255
4   beq r2, zero, no_safeproc # test if response was zero
5
6   safeproc: # it was not zero: we were accepted to safe-processing
7   call SAFECODE_START # Run safe code ...
```

Figure 7. Code Snippet of IRQ Handling a Lockstep Processing request

The corresponding `lock_step_monitor` architecture (Figure 8) consists of three subcomponents, of which only the `voter` component is safety critical.

The `controller` component is used to trigger safe-processing via the `request_sp` input from an external entity or via the `control_bus`. The number of participants, which must be an odd number greater or equal to three, so that the `voter` always has a majority and no ties, can also be set over the `control` bus. Once a safe-processing operation has been requested, the `irq` output, used to request more `processing_blocks`, will be asserted until notified by the `lockstep_processing` signal from the `synchronizer` component that lockstep processing has begun.

The implementation of the `synchronizer` component is kept simple. After detection of an initial read request all responding `processing_blocks` will be stalled until at least N `processing_blocks` have issued a read request on `psyb`. When this occurs the first N responders will be selected and their read transactions will be answered positively. At the same time, the selected `processing_blocks` will have their bit set in the enabled signal vector (`enabled[1..N]`.) A `processing_block` will read a negative result from its read transaction if it is surplus to requirements and may return to normal processing. Systematic errors could be avoided by the random choice of `processing_block`, this is left to later work.

The `voter` component is composed of three sub-components, the `compare_matrix` compares each of the n bus inputs (`plsb`) to every other bus input and exposes an n×n Boolean matrix as output interface. The `majority_voter` takes the output of the `compare_matrix` and selects an input representing the majority result for the `bus_multiplexer`, the `synchronizer` will tell the `majority_voter` which inputs should be considered for voting. The `bus_multiplexer` takes all bus inputs and selects one for forwarding informed by the `majority_voter` which will select the first input that compares equal to at least M (out of N) inputs. If this criteria is not met then no input will be selected and the voted safe bus is kept idle.

The `observer` component is responsible for detection of availability errors. That means it measures the time between state transitions of the `synchronizer` and, in case of error, will signal such on the availability `error` line. This availability error is also asserted when the `majority_voter` reports that there is no valid majority.

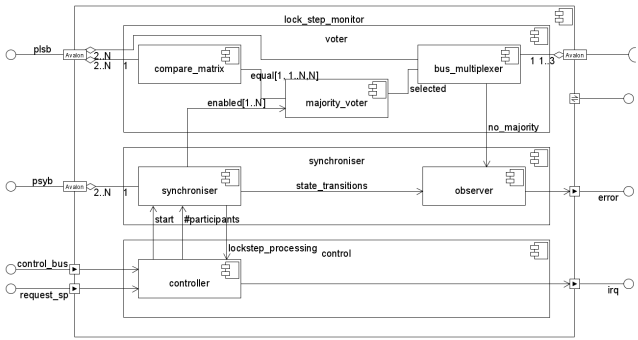

Figure 8: Architecture of the `lock_step_monitor`

The `lock_step_monitor` component was verified using the Open Verification Methodology on the cocotb platform. A demonstrator was built using a DE1-SoC board [12] which features an Intel Cyclone V FPGA; five NIOS II cores are instantiated to demonstrate dynamic lockstepping.

There are some caveats in the current design. The branch prediction in the individual NIOS II cores is dependent on execution history, which differs in each `processing_block`. This difference will result in additional latency in the execution of the `beq` instruction but can be mitigated by switching to static code prediction. We have ignored the effects of processor caches, by not using them, and we implemented a shadow stack, as the stack pointer also differs from processor to processor due to execution history. Configuring the NIOS with a shadow register set would help mitigating the effect of different stack-pointers on different processors executing the same code.

IV. CONCLUSION AND FURTHER WORK

We present a novel and promising proposal for dynamic lockstep operation of processors in multi-core/processor environments and some implementation notes for implementation in an FPGA. Future work includes investigating the optimal use of cache and expansion to lightly-coupled lockstepping.